\documentclass[useAMS,usenatbib]{mn2e}

\usepackage{graphicx}
\newif\ifpdf\ifx\pdfoutput\undefined\pdffalse\else\pdfoutput=1\pdftrue\fi
\newcommand{\pdfgraphics}{\ifpdf\DeclareGraphicsExtensions{.pdf,.jpg}\else\fi}

\defcitealias{LZ2001}{LZ01}
\defcitealias{Nandra1997b}{N97}

\newcommand{\D}{\displaystyle}

\title[Averaging the spectral shapes]{Averaging the spectral shapes}
\author[Piotr Lubi\'nski]{Piotr Lubi\'nski\thanks{E-mail: piotr@camk.edu.pl}\\
N. Copernicus Astronomical Center, Bartycka 18, 00-716 Warsaw, Poland}

\begin{document}
\pdfgraphics

\date{Accepted ... . Received ...; in original form ...}

\pagerange{\pageref{firstpage}--\pageref{lastpage}} \pubyear{2003}

\maketitle

\label{firstpage}

\begin{abstract}
The methods of obtaining the average spectral shape in a low statistics
regime are presented. Different approaches to averaging are extensively tested 
with simulated spectra, based on the $ASCA$ responses. The issue of binning up
the spectrum before fitting is discussed together with the choice of statistic
used to model the spectral shape. The best results are obtained with methods 
in which input data are represented by probability density functions. 
Application of weights, representing the coverage between the input and output 
bins, slightly improves the resolution of averaging.
\end{abstract}


\section{Introduction}

Spectral information obtained from X-ray astronomy instruments is usually the 
result of a compromise between the aim of achieving the best possible 
energy/spatial resolution and the purpose of collecting a large number of 
photons. In consequence, there will be always some class of objects which are
too faint, with spectra which cannot be studied with all desired particulars. 
The problem of a lack of statistics can be solved to some extent by averaging 
a number of weak spectra, but this implies examination of rather common 
properties. Nevertheless, since spectra can be grouped into subsamples 
according to some better established quantities such as the continuum slope, 
flux, hardness ratio, etc., this method may be quite powerful for studying 
various correlations.

A clear distinction should be made  between the average spectrum and the 
average spectral shape. The former is simply the average flux and the result 
is dominated by the brightest objects or states (in the case of average for 
a single object). The latter is the average of a relative quantity, spectral 
shape, usually defined as the ratio between data and a simple continuum model, 
common for all studied objects. Such a proportion is often employed to bring 
some discrete features into prominence --- a well known example is fig. 1 of 
\citet{Tanaka1995} where the redshifted iron K$\alpha$ line profile, observed 
for MCG-6-30-15, was shown. The shape defined as above is customarily used only
for illustrative purposes, e.g., \citet{Reynolds1997} and \citet{Reeves2003}. 
However, the average shape can be constructed and studied in a quantitative way, 
as was done for the spectra of Sy1 nuclei observed by $ASCA$ \citep{Nandra1997a,
Nandra1997b}. A similar investigation was performed later for a larger sample 
of Sy1 $ASCA$ observations \citep{LZ2001}, where average shape spectra were 
obtained for subsamples grouped according to the continuum slope. The 'average 
shape spectrum' is defined as the average data to continuum model ratio (i.e.,
the average shape), multiplied by the average continuum model (model with the 
average parameters calculated with a standard weighted mean).

The average profiles of the iron K$\alpha$ line presented in \citet{Nandra1997a} 
and in \citet{LZ2001} are clearly distinct. This difference was ascribed 
partially to the change in calibration of $ASCA$ SIS detectors done after 1997 
and partially to a different approach to averaging. However, the first 
explanation was later questioned and the whole difference was assigned to the 
dissimilar averaging procedure \citep{Yaqoob2002}. The issue of changed 
calibration will be discussed elsewhere; here we want to consider the problem 
of correct averaging. Additional motivation comes from the fact that it seems 
promising to apply similar averaging procedures to the data from other missions, 
such as $Chandra$, $XMM$--$Newton$, and, in future, $Astro$--$E$. Therefore, it is 
important to have at one's disposal a verified method, exploiting all available 
information in the most efficient and accurate way.

In the following sections we discuss three basic aspects of spectral shape 
averaging: bin weights, prebinning and the character of the data. The first is 
the way in which the information on the relative positions of the input and 
output bins is taken into account. Prebinning means here the summing of counts 
from a single input spectrum over the span of the output bin. Finally, the
Poisson character of the data is important for low numbers of counts where the 
standard weighted average, which assumes a symmetric, Gaussian probability 
density distribution, cannot be used. These aspects are connected --- for 
example, prebinning leads to the loss of some information on the data 
distribution in input bins, but, on the other hand, increases the number of 
counts. Various approaches to these basic issues can be combined to construct 
different averaging methods; our aim is to test them through extensive tests 
performed on simulated data.
 
\section{Averaging method}

\subsection{Rebinning}
\label{rebin}

By definition, the average value of a function $\xi(e)$ over the range of its
argument, $(e_{i},e_{i+1})$, is equal to the ratio of this function integral 
over this range to the length of this range, $\Delta e_{i}$,

\begin{equation}
\langle \xi \rangle _{i} = \frac{\int_{e_{i}}^{e_{i+1}} \xi(e) de}
{\int_{e_{i}}^{e_{i+1}} de} 
= \frac{\int_{e_{i}}^{e_{i+1}} \xi(e) de}{\Delta e_{i}}.
\label{adefinition}
\end{equation}

Here and after we use the following convention: angle brackets $\langle 
\rangle$ denote the average over some range of the function argument, i.e., 
rebinned value, without weights associated with the accuracy of the averaged
quantity. Any average, for a single spectrum or for many spectra, weighted by 
the accuracy weights is indicated by a dash over the symbol. Subscripts $i, j, 
k$ are used for input data, spectra and output data, respectively. To 
distinguish if the input data are summed or averaged for a single spectrum or 
for all spectra, we will use different upper summation limits, $n_{kj}$ for 
a single, $j$-th spectrum and $n_{ks}$ for all spectra. The output data for 
an individual spectrum will be denoted by an additional index $j$. Finally, 
the number of averaged spectra is equal to $n_{s}$.

\begin{figure}
\begin{center}
\includegraphics[width=\columnwidth]{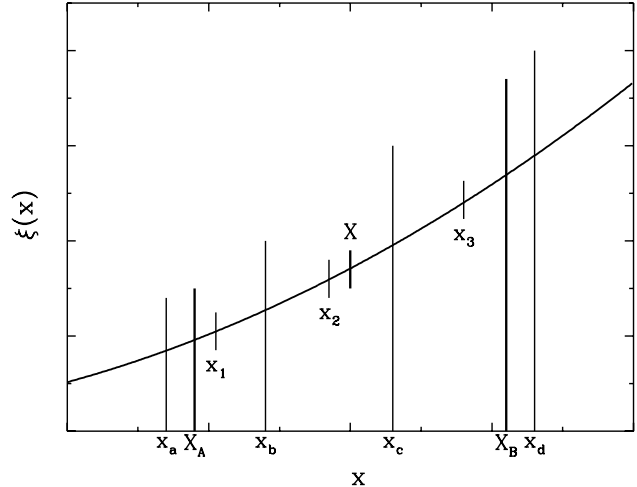}
\end{center}
\caption{\small Idea of rebinning of function over some range of its argument. 
The average for segment $(E_{k},E_{k+1})$ is determined using the known averages 
for segments $(e_{0},e_{1})$, $(e_{1},e_{2})$, ..., $(e_{n_{kj}-1},e_{n_{kj}})$.}
\label{rebini}
\end{figure}

Assume that we know the averages of a certain energy function $\xi$ for some 
initial distribution of energy ranges {$\Delta e_{i}$} and we want to determine 
averages for another set of energy ranges {$\Delta E_{k}$}. This is rebinning: 
for a given spectrum, values in some bins are converted to values in 
bins occupying different ranges. The idea of rebinning is illustrated in 
Fig.~\ref{rebini}. In general, the output bin $k$, with boundaries $(E_{k},
E_{k+1})$, for a spectrum numbered with $j$ expands over $n_{kj}$ input bins 
(with lower limits $e_{0},e_{1},...,e_{n_{kj}-1}$ and upper limits 
$e_{1},e_{2},...,e_{n_{kj}}$). The first and the last input bins can lie 
partially outside the output bin. Input data are discrete, we know only the 
average of the unknown function $\xi(e)$ measured by the detector over the 
$i$-th bin, its value $\langle \xi \rangle _{i}$ is attributed to the centre 
of the bin. Using the approximation that $\xi(e)$ is constant within an input 
bin and introducing bin weights $b_{i}$ we obtain the formula for the rebinned 
value $\langle \xi \rangle _{kj}$ in the form

\begin{equation}
\langle \xi \rangle _{kj} = \sum_{i=1}^{n_{kj}} b_{i} \langle \xi \rangle _{i},
\label{rebinned}
\end{equation}

where

\begin{equation}
b_{i} = \cases{\frac{\D e_{1}-E_{k}}{\D \Delta E_{k}},
& {\small partial overlap, left boundary;}
\cr \frac{\D \Delta e_{i}}{\D \Delta E_{k}},
& {\small full overlap, inside output bin;}
\cr \frac{\D E_{k+1}-e_{n_{kj}-1}}{\D \Delta E_{k}},
& {\small partial overlap, right boundary;}
\cr 1,
& {\small input bin covers output bin,}}
\end{equation}

and

\begin{equation}
\sum_{i=1}^{n_{kj}} b_{i} = 1.
\end{equation}

Defining $\delta _{i}$ as the width of the overlap between input bin $i$ and 
output bin $k$, the weights $b_{i}$ may be simply expressed as  
$\delta _{i}/\Delta E_{k}$.

Boundary input bins, with the same value of weight $b_{i}$, can occupy quite
different ranges outside the output bin. Hence, they can represent different
information on the averaged function, integrated over a different range of
argument. To take this fact into account one can apply another bin weight,
inversely proportional to the input bin width, equal to $1/\Delta e_{i}$ or,
better, to $b_{i}/\Delta e_{i}$.

\begin{figure}
\begin{center}
\includegraphics[width=\columnwidth]{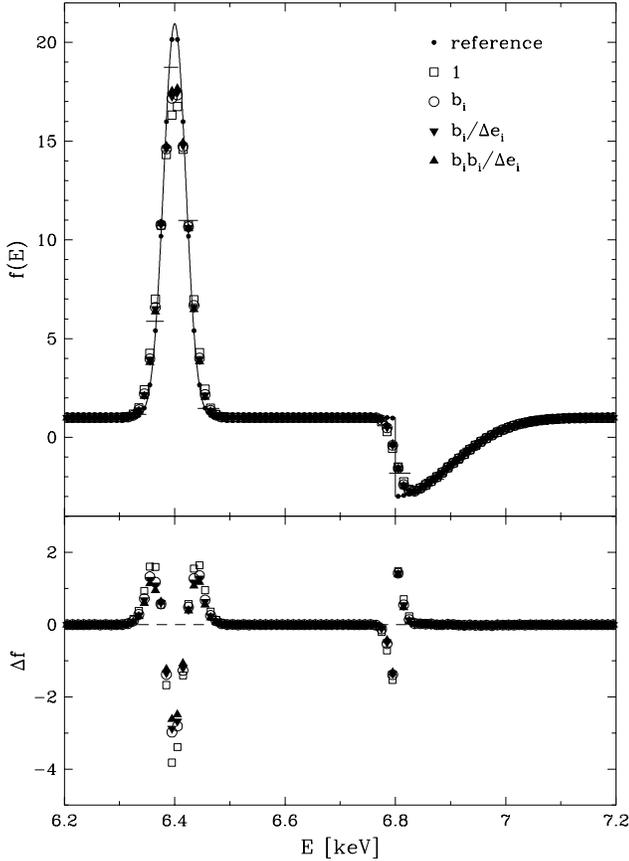}
\end{center}
\caption{\small Upper panel: Averaging of function with and without bin 
weights. Solid line shows the averaged function f(E), reference shape (dots) is 
the result of integrating this function over 0.01 keV bins. Averaged data are 
obtained by integrating f(E) with random bin widths taken from 0.01-0.02 keV 
range (an example of these bins is shown with the horizontal lines). 
The average shapes were calculated for 200 input data sets, with output bin 
widths equal to 0.01 keV and using 4 different bin weights. Lower panel shows 
the differences between the results of averaging and the reference shape.}
\label{baveran}
\end{figure}

Figure \ref{baveran} shows an example of averaging with and without bin 
weights. The averaged function $f(E)$ is a constant plus Gaussian peak 
plus an edge, modelled by a negative half-Gaussian. The widths of 
Gaussians were set to 0.02 keV and 0.1 keV for peak and edge, respectively.
This function was integrated over 0.01 keV bins to obtain the reference shape 
after using the instrument with 0.01 keV energy resolution. The function was 
then integrated 200 times for random binning patterns, with bin widths $\Delta 
e_{i}$ taken from some interval. The resulting 'spectra' were rebinned in four 
different ways (with bin weights listed in Fig. \ref{baveran}) to 0.01 keV 
output bins and the results were compared with the reference shape. 
The whole procedure was repeated for different ranges of input bin widths. 
When the input bins $\Delta e_{i}$ are much broader than 0.01 keV the resulting 
averages with and without bin weights are indistinguishable. For input bin 
widths comparable to or less than 0.01 keV, the differences in results correlate 
with decreasing input bin width. Procedures with bin weights clearly better 
reproduce the sharp features of the averaged function than the simple 
arithmetic average, the best of these is the weight in the form $b_{i}^{2} 
/\Delta e_{i}$ but the differences are rather small.

Bin weights play a role similar to approximating the shape of the rebinned 
function with a polynomial spline. However, for poor quality data they are 
safer than any iterative spline procedure, because the latter method may 
amplify some spurious spectral features during consecutive iterations. 
Therefore, since we have to work with low statistics data and since the energy 
resolution of $ASCA$ SIS detectors is not so high, the discrete spectral 
features can be well traced by bin weights alone, without using an additional 
spline approximation. Moreover, any more complex spline procedure applied to 
a small set of weak spectra may lead to a quite accidental approximation of the 
local spectral shape. On the contrary, bin weight values are well defined, do 
not depend on the averaged function value and control the relevance of 
information given by the input bin, comparing simply input and output bin 
locations.

\subsection{Prebinning}
\label{prebinning}

The quantity which we intend to average is the ratio of measured and modelled
fluxes. Since the flux is proportional to the number of counts we can utilize
the fact that the number of counts measured for a broader bin is equal to the 
sum of counts collected for narrower bins constituting this broad bin. 
Then, instead of averaging input flux ratios for a given output bin, we can 
directly determine data and model fluxes for that bin. The measured flux, 
$\langle f \rangle _{kj}$, is equal to the sum of all net counts $D_{i}$, 
divided by appropriate detector area $A_{i}$, and normalized with the 
observation time, $T$, and output bin width, $\Delta E_{k}$, 

\begin{equation}
\langle f \rangle _{kj} = \frac{\sum_{i=1}^{n_{kj}} g_{i} D_{i}/A_{i}}
{T\Delta E_{k}},
\label{sumflux}
\end{equation}

where weights $g_{i}=\delta_{i}/\Delta e_{i}$ ($=b_{i}\Delta E_{k}/\Delta 
e_{i}$) are introduced for boundary input bins, only partially overlapping 
with the output bin.

In practice, there is no simple method of performing spectral fitting with 
arbitrary energy bins, i.e., with energy bins adjusted to cover the output 
bins. This is due to the fact that the instrumental response function is 
defined as a matrix for a fixed set of energy bins. Therefore, it is necessary 
to perform prebinning with model counts determined for input channels instead 
of fitting them for output bins. The modelled flux, $\langle f \rangle ^{m} 
_{kj}$, is expressed through the model net counts, $M_{i}$,

\begin{equation}
\langle f \rangle ^{m} _{kj} = \frac{\sum_{i=1}^{n_{kj}} g_{i} M_{i}/A_{i}}
{T\Delta E_{k}},
\label{sumfluxm}
\end{equation}

The numerator in (\ref{sumflux}) can be replaced by $\sum_{i=1}^{n_{kj}} g_{i} 
D_{i}/A_{k}$, where the area $A_{k}$ represents the effective detector 
efficiency for bin $k$. After similar replacement in (\ref{sumfluxm}), 
the ratio of the data and model fluxes for output bin $k$ is equal to

\begin{equation}
\langle r \rangle _{kj} = \frac{\sum_{i=1}^{n_{kj}} g_{i} D_{i}}
{\sum_{i=1}^{n_{kj}} g_{i} M_{i}}.
\label{prebin} 
\end{equation}

The procedure with prebinning is simpler than that based on averaging all input 
ratios $\langle r \rangle _{i}$ (=$D_{i}/M_{i}$); after prebinning one has only
to average ratios $\langle r \rangle _{kj}$ obtained for the output bins from 
various spectra. Nevertheless, there is one disadvantage: the information 
included in the input ratios $\langle r \rangle _{i}$ is lost. The issue of
how this affects the results by reducing the resolution will be discussed 
later, in Sec. \ref{discuss}.

\subsection{Accuracy weights}

The initial ratios $\langle r \rangle _{i}$ are measured with some finite 
accuracy and this should be taken into account in averaging. For the set 
of $n_{ks}$ independent ratios, described by the probability 
density functions $p_{i}(r)$, the mean value (centre of gravity) is given 
by the integral

\begin{equation}
\overline r_{k} = \frac{\int_{-\infty}^{\infty} r p_{k}(r) dr}
{\int_{-\infty}^{\infty} p_{k}(r) dr},
\label{intmean}
\end{equation}

where the joint density function $p_{k}(r)$ is equal to the product of 
partial densities

\begin{equation}
p_{k}(r) = \prod_{i=1}^{n_{ks}} p_{i}(r).
\end{equation}

The standard deviation for  $\overline r_{k}$, $\sigma _{\overline r_{k}}$, 
is calculated from the variance definition

\begin{equation}
    \sigma _{\overline r_{k}}^{2} = \frac{\int_{-\infty}^{\infty} 
    (r-\overline r_{k})^{2}p_{k}(r)dr}{\int_{-\infty}^{\infty} p_{k}(r)dr}.
\label{variance}
\end{equation}

\subsection{Standard average}

In the case where the densities $p_{i}(r)$ have Gaussian shape, centered at
$\langle r \rangle _{i}$ and with width parameters $\Delta \langle r \rangle 
_{i}$, one obtains from (\ref{intmean}) and (\ref{variance}) the standard 
formula for the weighted mean and its uncertainty

\begin{equation}
\overline{r} _{k} = \frac{\sum_{i=1}^{n_{ks}} w_{i} \langle r \rangle _{i}}
{\sum_{i=1}^{n_{ks}} w_{i}},
\quad \quad \Delta \overline{r} _{k} = \frac{1}
{\sqrt{\sum_{i=1}^{n_{ks}} w_{i}}},
\label{weighted}
\end{equation}

where weights $w_{i}$ are equal to $1/(\Delta \langle r \rangle _{i})^{2}$. 

The above formula can be applied to ratios obtained from larger numbers of 
counts, when the Poisson probability function associated with the data can 
be approximated by a Gaussian function. In such a case, the unknown 
uncertainty of the true number of counts can be approximated by the square 
root of the measured number of counts. Then, the ratio uncertainty $\Delta 
\langle r \rangle _{i}$ is equal to $\sqrt{g_{i}N_{i}+g_{i}B_{i}}/g_{i}M_{i}$, 
where $N_{i}$ denotes the source (net effect + background) number of counts 
and $B_{i}$ is the background number of counts. Weights $g_{i}$ modify these 
numbers for an input bin lying on the boundary of the output bin. In Appendix 
\ref{appena} it is shown that for relative quantities such as flux, the 
accuracy weights $w_{i}$ to some extent play a role similar to the bin weights. 

Obviously, the condition of large number of counts can be more easily fulfilled 
in the case of prebinned data, when the output bins are broader than the input 
ones. The uncertainty of ratio $\langle r \rangle _{kj}$, given by equation 
(\ref{prebin}), is then approximated by

\begin{equation}
\Delta \langle r \rangle _{kj} = \frac{\sqrt{\sum_{i=1}^{n_{kj}} g_{i} 
N_{i}+\sum_{i=1}^{n_{kj}} g_{i}B_{i}}}{\sum_{i=1}^{n_{kj}} g_{i} M_{i}}.
\label{prebinned}
\end{equation}

The final ratio $\overline{r}_{k}$ and its uncertainty $\Delta 
\overline{r}_{k}$ for prebinned data, averaged over all spectra, is calculated 
from (\ref{weighted}), with $\langle r \rangle _{i}$ replaced by 
$\langle r \rangle _{kj}$, weights $w_{kj}$ equal to $1/(\Delta \langle 
r \rangle _{kj})^{2}$ and summation going from $j$ = 1 to $j$ = $n_{s}$.

\subsection{Combined weights}
\label{combined}

According to the results of Sec. \ref{rebin}, the accuracy weighted average 
for input ratios $\langle r \rangle _{i}$ should be modified to incorporate 
bin weights. Application of another type of weight, such as bin weights $b_{i}$, 
can be realized via broadening of the probability density distributions 
associated with the data by raising them to a power equal (or proportional) to 
this additional weight. It should be stressed that this procedure is used only 
to change the $relative$ widths of these distributions, by taking into account 
the overlap between given input bin $i$ and output bin $k$. There is, then, 
some arbitrariness in defining the broadening power index. In tests with 
simulated spectra, we found that the change of $b_{i}$ to $b_{i}/10$ or 
$10b_{i}$ affects only the fifth digit in the result of averaging. Nevertheless, 
any broadening of the initial distributions leads to a change of the width of  
resulting distribution, hence affects the error of $\overline{r} _{k}$. For 
narrow input bins ($b_{i} \ll 1$) the final distribution can be quite broad and 
its dispersion should be renormalized, since bin weights are used only to 
redefine the calculation of the average, and should not change the accuracy of 
the input $\langle r \rangle _{i}$ values. To avoid absolute narrowing of any 
of initial distributions the renormalization is done by replacing $b_{i}$ in 
the above equation by $\hat{b}_{i} = b_{i}/max(b_{i})$. This leads to only 
slightly broader distributions than those obtained without bin weights.

The combined weights for the standard average are equal to $\hat{b}_{i}w_{i}$, 
since the width of broadened Gaussian is equal to $\sigma _{i} /\sqrt{\hat{b}
_{i}}$.  However, the combined weighted average and its propagated error, 
calculated from (\ref{weighted}), can be expressed using weights $b_{i}$ 
directly, due to the fact that the normalizing factors $1/max(b_{i})$ in 
numerator and denominator cancel out 

\begin{equation}
\overline{r} _{k} = \frac{\sum_{i=1}^{n_{ks}} b_{i} w_{i} r _{i}}
{\sum_{i=1}^{n_{ks}} b_{i}w_{i}}, \quad \quad
\Delta \overline{r} _{k} = \frac{\sqrt{\sum_{i=1}^{n_{ks}} b_{i}^{2} 
w_{i}}}{\sum_{i=1}^{n_{ks}} b_{i} w_{i}}.
\label{incorrect}
\end{equation}

The above formulae were used to obtain the average spectral shape for Sy1 
active nuclei observed by $ASCA$ \citep{LZ2001}. However, some correction
should be made: weights $w_{i}$ for border input bins should be decreased by 
a factor equal to weight $g_{i}$, since the $\langle r \rangle _{i}$ error 
is increased by $\sqrt{1/g_{i}}$ due to only a partial share of input bin 
flux in the flux of the output bin. In consequence, we obtain

\begin{eqnarray}
\overline{r} _{k} = \frac{\sum_{i=1}^{n_{ks}}b_{i}g_{i}w_{i} r_{i}}
{\sum_{i=1}^{n_{ks}} b_{i}g_{i}w_{i}}, \quad \quad
\Delta \overline{r} _{ks} = 
\frac{\sqrt{\sum_{i=1}^{n_{ks}} b_{i}^{2}g_{i}w_{i}}}
{\sum_{i=1}^{n_{ks}} b_{i}g_{i}w_{i}}.
\label{correct}
\end{eqnarray}

As discussed in Sec.~\ref{dibiased}, the above correction affects mainly
the ratio errors, leaving the ratio values almost unchanged.

\subsection{Average for Poisson data}
\label{data}

The formula given by (\ref{weighted}) is derived from the definition of the
mean (Eq. \ref{intmean}) for quantities described by the probability 
density function in  Gaussian form. However, the numbers of counts collected in 
a single channel by SIS instruments of $ASCA$ for weak sources such as AGNs are 
very small --- above 7 keV these numbers are often equal to 1 or 0. Therefore,
equation (\ref{weighted}) cannot be used for a low number of counts, where the
Poisson distribution differs substantially from the Gaussian one. Moreover, 
in this situation the unknown uncertainty of the true number of counts cannot 
be approximated by the square root of the measured number of counts. 

Individual densities in the counts space, $p_{i}(\lambda)$, including
the case of border input bins with numbers of counts modified by weights 
$g_{i}$, can be expressed as Poisson functions of unknown mean $\lambda$, 
with observed $g_{i}N_{i}$ source counts and $g_{i}B_{i}$ background counts 
as parameters

\begin{equation}
    p_{i}(\lambda) = C_{i}\frac{e^{-(\lambda+g_{i}B_{i})}(\lambda+g_{i}
    B_{i})^{g_{i}N_{i}}}{\Gamma(g_{i}N_{i}+1)},
\label{poiss}
\end{equation}

where constant $C_{i}$ ensures proper normalization of the probability
function. The above formula has the form of a Poisson distribution, 
however, here the problem is inverted --- we are interested in a function of 
continuous argument $\lambda$ for given numbers of observed counts. 

Because numbers of counts modified by weights $g_{i}$ are not necessarily the 
integer numbers, the standard factorial present in the Poisson distribution 
definition, in Eq. (\ref{poiss}) is replaced by the complete gamma function 
$\Gamma(g_{i}N_{i}+1)$. For the same reason we cannot give an analytical 
expression for the normalizing factor, (cf. Eq.~(\ref{appois}) in Appendix 
\ref{appenb}), because for non-integer $g_{i}N_{i}$ the expansion of 
$(\lambda+g_{i}B_{i})^{g_{i}N_{i}}$ binomial is not finite, and value of 
$C_{i}$ has to be computed by numerical integration.

The formula (\ref{poiss}) is based on the assumption that the background number
of counts is known precisely. Evidently, in a real situation we do not known
the true background rate. However, as will be shown in Sec. \ref{discuss},
this assumption is sufficient to give proper averaging results. Our main
goal is to describe the data probability distribution in a way more valid than 
a Gaussian distribution, and Eq. (\ref{poiss}) leads to a satisfactory 
result. 

Nevertheless, in Appendix \ref{appenb} we present two solutions to the problem 
of the unknown background which we have found in literature. As can be 
expected, these procedures lead to a more diffuse distribution than that given 
by (\ref{poiss}), especially with a broadened left tail when the observed 
background number of counts is close to zero. We have tried to include these 
methods in the averaging code, but, because of some problems with the
implementation of these methods in current versions of popular spectral fitting 
codes (see next Section) we cannot model the continuum in a corresponding way.
Therefore, the treatment of data in averaging and in continuum modelling cannot 
be consistent, and, in consequence, the data/model ratios will diverge.
 
Direct calculations based on Eq. (\ref{poiss}) are impractical for larger 
values of $g_{i}N_{i}$. Therefore the function $p_{i}(\lambda)$ is calculated 
using the algorithm invented for computing the binomial distribution 
\citep{Loader2000}. Computation, except for the trivial case when $\lambda+g_{i
}B_{i}$ = 1, reduces to calculating the exponent of some function $\alpha _{i}
(\lambda)$, depending on $g_{i}$ and $B_{i}$, normalized with some factor 
$\beta _{i}$ depending on $N_{i}$, $B_{i}$ and $g_{i}$. Then the probability 
density can be given in the form

\begin{equation}
    p_{i}(\lambda) = \frac{e^{\alpha _{i}(\lambda)}}
    {\beta _{i}(g_{i}N_{i},g_{i}B_{i})}.
\end{equation}

Turning to ratios, we take into account the modelled number of counts, 
$g_{i}M_{i}$, and the averaging reduces to determining the mean of the 
joint density distribution function

\begin{equation}
    p_{k}(r) = \prod_{i=1}^{n_{ks}} C_{i}e^{\alpha' _{i}(r)},
\label{probin}
\end{equation}

where the function $\alpha' _{i}$ is equal to $\alpha _{i}$ with an argument 
scaled by the factor $1/g_{i}M_{i}$ and the  normalizing factors $C_{i}$ are 
functions of $N_{i},B_{i}$, $M_{i}$ and $g_{i}$. 

In the case of prebinned data, for a spectrum numbered with $j$ we have
function $\alpha' _{kj}$ and factorial $C_{kj}$ depending on the summed counts
$\sum_{i=1}^{n_{kj}}g_{i}N_{i}$, $\sum_{i=1}^{n_{kj}}g_{i}B_{i}$ and 
$\sum_{i=1}^{n_{kj}}g_{i}M_{i}$, and the joint density distribution function 
is calculated for $n_{s}$ spectra

\begin{equation}
    p_{k}(r) = \prod_{j=1}^{n_{s}} C_{kj}e^{\alpha' _{kj}(r)}.
\label{proprebin}
\end{equation}

At last, when bin weights $\hat{b}_{i}$ are taken into account for data without
prebinning, the formula for the joint probability density functions has the 
form

\begin{equation}
    p_{k}(r) = \prod_{i=1}^{n_{ks}} \Big [C_{i}e^{\alpha' _{i}(r)}
    \Big ]^{\hat{b}_{i}}.
\label{prorebin}    
\end{equation}

\begin{figure}
\begin{center}
\includegraphics[width=\columnwidth]{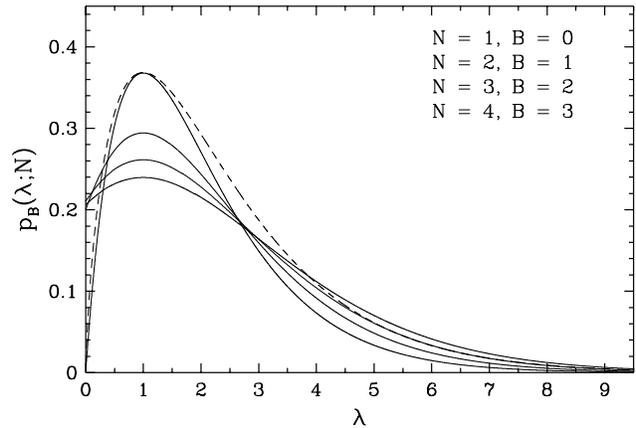}
\end{center}
\caption{\small Four probability functions (Eq. (\ref{poiss})) for net counts 
D~= 1 and background counts B equal to 0,1,2,3 (from top). For N = 1 the 
example of broadening of the probability density function by the bin 
weight = 0.75 is shown with a dashed line.}
\label{rewo}
\end{figure}

There is no computer on Earth able to directly multiply many very small 
numbers, which are unavoidable for a broad range of ratio values, especially 
when $n_{ks}$ is of the order of thousand and the multiplied distributions are 
narrow. Thus it is necessary to replace $p_{k}(r)$ with its logarithm

\begin{eqnarray}
\nonumber
    \log p_{k}(r) & = & \log \prod_{i=1}^{n_{ks}} 
    \Big [C_{i}e^{\alpha' _{i}(r)}\Big ]^{\hat{b}_{i}} 
    = \sum_{i=1}^{n_{ks}} \log \Big [C_{i}e^{\alpha' _{i}(r)}\Big ]
    ^{\hat{b}_{i}} = \\
    & = & \sum_{i=1}^{n_{ks}} \hat{b}_{i}\alpha' _{i}(r)
    +\sum_{i=1}^{n_{ks}}\hat{b}_{i} \log C_{i}.
\end{eqnarray}

The second sum in the last line of the above equation can be dropped since it 
only corresponds to adding a constant to the density distribution function 
and the computed density function is 

\begin{equation}
    \log p_{k}(r) = \sum_{i=1}^{n_{ks}} \hat{b}_{i}\alpha' _{i}(r).
\end{equation}

\begin{figure}
\begin{center}
\includegraphics[width=\columnwidth]{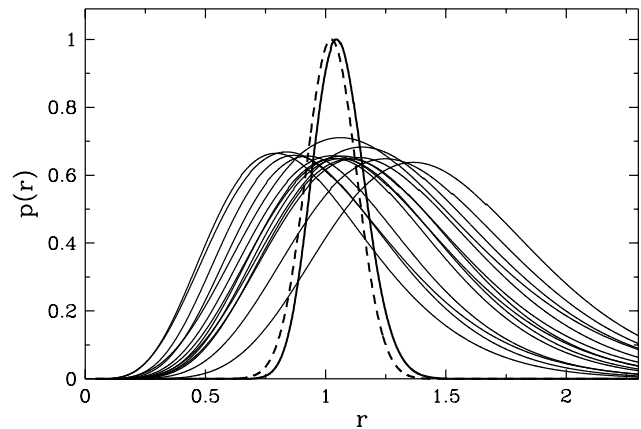}
\end{center}
\caption{\small Example of the joint probability distribution applied to 
determine the mean ratio. Thin lines show the Poisson probability functions 
representing the averaged input ratios, thick solid line shows their product 
distribution, normalized to have maximum equal to 1. Dashed line illustrates
the product distribution obtained when the input data are given by the Gaussian 
probability functions.}
\label{gamma}
\end{figure}

The final average $\overline r_{k}$ is found from equation (\ref{intmean})
as the mean of the exponent of the above function but with the lower integral 
limit equal to 0, since the number of counts (or ratio) cannot be negative. 
Accordingly, the accuracy of $\overline r_{k}$ is calculated from  Eq. 
(\ref{variance}). 

The joint density function $p_{k}(r)$ is constructed as a likelihood function 
in the maximum likelihood estimation method. Then, the best estimator for 
$\overline r_{k}$ is the mode, for which the accuracy should be calculated
using the second derivative of $p_{k}(r)$. The computation of mode for
density function given by Eq. (\ref{intmean}) is simple but numerical
determination of its second derivative and then its expectation value to
assign the accuracy to $\overline r_{k}$ for many input bins may
significantly increase the computation time. Therefore, because the final 
distribution functions are usually quite symmetric, especially for a larger 
number of averaged spectra, the mode and its error were replaced here by the 
mean and standard deviation. As it was tested, even for an extreme case of a
single, weak spectrum, the difference between mode and mean of $p_{k}(r)$ does 
not exceed 0.5\%.

\section{Continuum model}

Considering the issue of averaging the data/model ratios in 
Sec.~\ref{prebinning}, we have concluded that the continuum model should be 
fitted to the non-binned spectrum, i.e., collected with original, single SIS 
channels. Due to this fact, for a low number of counts we also have to treat the 
spectral modelling in a non-standard manner. Since the $\chi^{2}$-statistic 
cannot be applied during fitting, an alternative approach is needed. A fairly 
popular and well established solution is the C-statistic \citep{Cash1979}. It 
is based on the maximum likelihood method, and has no limitation according to 
the number of counts. In the case of $N_{i}$ source counts and $S_{i}$ model 
counts it is defined as \citep{FDS2001}

\begin{equation}
C = 2\sum_{i=1}^{n_{c}}(S_{i}-N_{i}\ln S_{i}),
\end{equation}

where $n_{c}$ is the number of data channels.

The difference of two Poisson distributions is not Poisson distributed, thus
the C-statistic cannot be applied to background subtracted data. Therefore in 
the C-statistic case the source and background spectra were fitted 
simultaneously. The net number of model counts $M_{i}$ was determined by 
subtracting the appropriate numbers modelled for a given channel. In 
consequence, data in modelling were treated in a way consistent with averaging 
based on equation (\ref{poiss}). 

The C-statistic is implemented in both most popular X-ray spectra fitting codes,
XSPEC \citep{Arnaud1996} and SHERPA \citep{FDS2001}. However, used within
XSPEC it produced biased results: the fitted power law indices and 
normalizations were much larger than those assumed in the simulated 
model.\footnote{There was a bug in XSPEC, repaired in the
its version 11.2.0bs, after preparation of this paper. Now XSPEC used with 
the C-statistic produces results consistent with those obtained with SHERPA.}
Therefore, in the case of non-rebinned spectra, fitting the model of continuum 
used to normalize observed data to obtain the spectral shape was done only 
with SHERPA. For comparison, we have also binned  up spectra and fitted them 
with a $\chi^{2}$-statistic using both XSPEC and SHERPA. All tests were 
performed with XSPEC, version 11.1 and SHERPA, version 2.3.

There is a Bayes statistic option in SHERPA, which corresponds to the method
derived by \citet{Loredo1992}. We have tested also this approach during 
spectral fitting, unfortunately the results were similar to the results
obtained with the C-statistic and XSPEC: the fitted power law was much steeper
than the model assumed in the simulation of the spectrum. Hence, since the 
usage of the Bayes statistic in SHERPA seems to be somewhat uncertain and since
the results obtained with C-statistic and SHERPA are quite satisfying, we
do not use the Bayes approach in the tests described later.

All continuum models for the source and background spectra were fitted in the 
energy range (3-4.5,7.5-10) keV, i.e., the reference continuum shape was 
determined in the vicinity of the Fe $K_{\alpha}$ line but without the line 
region itself. We used a power law model, assuming independent slope
parameters for source and background and independent normalization for each
spectrum.

\section{Tests}

\subsection{Reference shape}
\label{shape}

The procedure developed for averaging the spectral shapes has to be verified, 
moreover, some tests of its alternatives are needed. It is obvious that such 
tests cannot be done for real data, as we must know the actual average shape 
to have a reference. For this purpose we have simulated 100 spectra for each of 
two SIS instruments. The reference model was similar to that obtained for the 
average $ASCA$ Sy1 nuclei spectral shape \citep{LZ2001}. Simulations were done 
with responses from different periods of $ASCA$ mission. The width of a single 
SIS channel was equal to 14.6 eV, since we used the BRIGHT2 data mode with
1024 channels. As the starting conditions we adopted exposure times and 
background spectra obtained for five observations of IC 4239A, which were short, 
with an elapsed time of 7-18 ks. The background spectra were extracted for 
rather small regions with a radius of 28.5 SIS pixels. Hence, we tested a 
rather extreme case of low statistics to be sure that the method behaves well 
on the boundary of its application area.

\begin{figure}
\begin{center}
\includegraphics[width=\columnwidth]{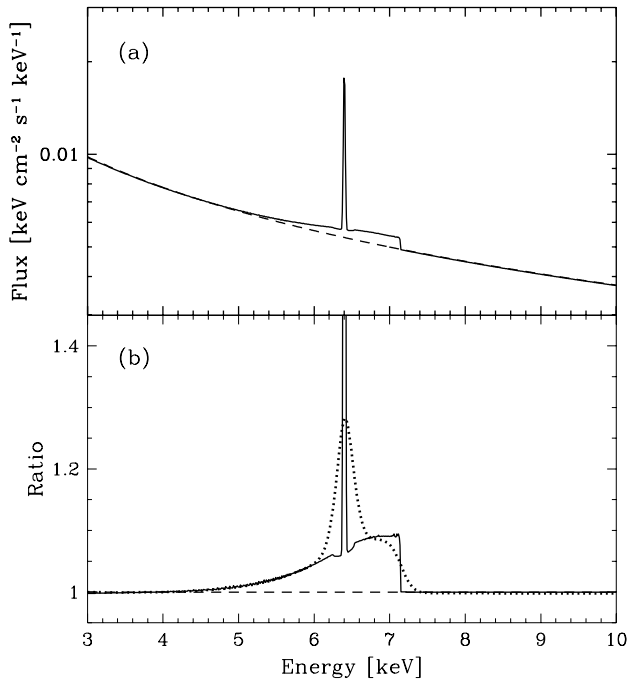}
\end{center}
\caption{\small (a) Model function (solid line) consisting of a disc line and 
a Gaussian together with the power law continuum. Dashed line shows the
continuum model, i.e., power law fitted outside the Fe line region, in the 
energy ranges 3-4.5 and 7.5-10 keV. (b) Ratio of the line and continuum model 
functions (solid line) compared with the data/model ratios (dots) obtained 
for the simulated reference spectra.}
\label{modelines}
\end{figure}

In the upper part of Figure \ref{modelines} we have shown our reference model,
consisting of a broad, disc line component, a narrow, Gaussian line component
and continuum in the form of power law. The parameters of the model were as
follows: power law index, $\Gamma=1.8$, power law normalization, $A=2.333
\times10^{-2}$ keV$^{-1}$cm$^{-2}$s$^{-1}$, disc line energy, $6.4$ keV, inner 
disc radius, $6\;GM/c^{2}$, outer disc radius, $1000\;GM/c^{2}$, disc emissivity 
in the form $(1-\sqrt{6/R})/R^{3}$, inclination, $45 \degr$, Gaussian line 
energy, $6.4$ keV, Gaussian line width, $0.01$ keV. The normalization of 
disc line and Gaussian components was adjusted to get equivalent widths equal 
to 130 eV and 60 eV, respectively.

Spectral shape is defined as the ratio between observed data and fitted 
continuum model, thus the reference model should be transformed in the same 
way. The reference data were simulated with a very long exposure time, 
$10^{9}$s, using responses of original IC 4329A spectra. Then, the continuum 
for the reference spectra was modelled with a power law model outside the iron 
line region. Resulting data/model ratios were averaged with weights equal to 
the original observation times, to take into account changes of SIS responses 
during satellite operation. These reference ratios are presented in the lower 
part of Fig. \ref{modelines}. Rebinning with finite bin width always leads to 
a change in the function shape, thus the results of averaging done with tested 
methods are compared not only with the reference ratios obtained for a single 
input bins, but also with the results of averaging these ratios to appropriate 
output bins. This reference averaging was done with our best method Ia 
(described below), but for such a high number of counts all procedures produce 
almost indistinguishable results.

\subsection{Tested methods}
\label{procedures}

Various averaging procedures can be constructed on the basis of three 
elements: accuracy weights, bin weights and prebinning. We have tested
some of these combinations to check their behaviour in different situations.
Table \ref{methods} presents the characteristics of the tested procedures. 
Methods numbered with I are based on determination of the joint density 
function, methods II are those using the accuracy weights $w_{i}$, and 
method III is the arithmetic average with prebinning. Among them, method IIe 
uses the standard weighted average (\ref{weighted}) applied to only those 
input bins, whose centres lie in the given output bin. Since this procedure 
is probably the one most commonly used, it will be termed 'standard' in the
rest of the paper.

Tested methods were applied to three types of 'data/model' data. The first
one was obtained for spectra with original SIS channels and the continuum
model fitted with the C-statistic. The data of the second type are those from 
spectra binned up\footnote{Regarding to arithmetics, 'binning up' is a special 
case of prebinning, introduced in Sec. 2.2, done for the ouput bin covering 
exactly the sum of input bins. Nonetheless, there is a clear difference in 
applying these two procedures, the former is used before fitting the continuum 
model, the second uses the results of the model fitted to the initial input 
bins.} firstly to gather at least 20 counts per channel, whereas 
continuum is modelled again with the C-statistic. The third is the case 
of binned up spectra and the model fitted with $\chi^{2}$ statistic.

The results of the tests are presented in Figs. \ref{arfig1}, \ref{arfig2},
\ref{arfig3}, the reference ratios are shown with a solid line, whereas the
reference average for output bin width equal to 0.1 keV is plotted with 
dots. Tests were performed for different widths of output bins, here we 
present only results for 0.1 keV bins since this value corresponds to the 
ASCA SIS resolution for the iron line energy.

In order to check how the fitted continuum models reproduce the reference model,
a weighted average of power law parameters was calculated for each data set.
These results are presented in Table \ref{meanmodels}. Since there is still
a small tail of the disc line component below 4.5 keV, the power law fitted to 
the reference data (simulated with long exposure) is slightly less steep 
($\Gamma = 1.798$) than the initial power law used in simulations 
($\Gamma = 1.8$). All models fitted with the C-statistic give almost correct 
results, whereas fitting with the $\chi^{2}$ statistic leads to a clearly 
biased result. A similar effect was already studied by the Chandra X-ray Center 
staff \citep{Freeman2001}. 

Non-uniform binning up of data before fitting changes the relative influence of 
different parts of the spectrum on the fitted model. For a simple continuum 
model like power law this effect is almost negligible (cf. results from row 2 
and rows 3,4 in Table \ref{meanmodels}), however, for more complex models such 
a procedure should be applied with some caution. 

\begin{table}
  \caption{Averaging methods tested with simulated spectra. PF denotes the 
  probability density function. Bin weights equal to 1 mean that the input
  bin is taken into account only in calculating the average for the output 
  bin containing its centre. }
  \begin{tabular}{@{}ccccc@{}}
  \hline
   Method & Accuracy & Bin     & Prebinning & Formula  \\
          & weights  & weights &            &            \\
  \hline
   Ia     &  PF      &  $g_{i}$  &  yes     & (\ref{proprebin}) \\
   Ib     &  PF      &  $g_{i}$  &  no      & (\ref{probin})    \\
   Ic     &  PF      &  $b_{i}g_{i}$  &  no & (\ref{prorebin})  \\
   IIa    &  $w_{i}$ &  $g_{i}$  &  yes     & (\ref{prebin},\ref
   {weighted},\ref{prebinned}) \\
   IIb    &  $w_{i}$ &  $g_{i}$  &  no      & (\ref{weighted})  \\
   IIc    &  $w_{i}$ &  $b_{i}g_{i}$  &  no & (\ref{correct})   \\
   IId    &  $w_{i}$ &  $b_{i}$  &  no      & (\ref{incorrect}) \\
   IIe    &  $w_{i}$ &  1  &  no            & (\ref{weighted})  \\
   III    &  1       &  $g_{i}$  &  yes     & (\ref{rebinned},\ref{prebin}) \\
\hline
\end{tabular}
\label{methods}
\end{table}

\subsection{Biased vs. unbiased binning up}
\label{biased}

The standard, initial binning up method, based on adding counts from single channels to 
get at least a given number of counts is biased in this sense 
that resulting bin widths are inversely proportional to the measured flux. 
Then, on average, broader input bins are more frequent for ratios below 1 than 
for ratios above 1. In consequence, bin weights $b_{i}$ are usually smaller for 
ratios $> 1$ than $b_{i}$ for ratios $< 1$ and this leads to a biased average 
shape. To test this effect we have prepared 'data/model' data with random 
binning, where binning up pattern for a given spectrum was taken from the 
other spectrum, randomly selected. In this way the binning should be 
non-biased, without correlation between bin widths and ratios. These spectra 
were fitted with C-statistic and they are the fourth type of 'data/model' data 
tested with some methods. 

\begin{table}
  \caption{Mean values of the continuum model parameters determined for
  various input data and fitting procedure combinations. The first row
  presents the results obtained for the reference data.}
  \begin{tabular}{@{}cccc@{}}
  \hline
   Binning up & Statistic & Normalization & Index \\
  \hline
   none   & both       & 23.28$\pm$0.01 & 1.798$\pm$0.001 \\
   none   & C, SHERPA  & 23.23$\pm$0.18 & 1.798$\pm$0.006 \\
   biased & C, SHERPA  & 23.12$\pm$0.19 & 1.794$\pm$0.006 \\
   random & C, SHERPA  & 23.26$\pm$0.19 & 1.800$\pm$0.006 \\
   biased & $\chi^{2}$, XSPEC & 23.55$\pm$0.22 & 1.829$\pm$0.007 \\
\hline
\end{tabular}
\label{meanmodels}
\end{table}

\subsection{Inhomogeneous shapes}
\label{inhomo}

Consider a special situation: there are two distinct classes of objects, with 
Fe line average shapes clearly different, and, in addition, objects of one 
class are much brighter than those of the second class. Then, if the spectra 
were taken in similar conditions (i.e., with approximately equal exposures), 
their weighted average will be far from the true, physical mean for these two 
classes. We have tested such a case by simulating 20 spectra with 10 times 
longer exposures and with a reference model different from the basic one, described in 
Sec. \ref{shape}. In this model all parameters are the same as in the basic 
one, only the disc line component is weaker, with equivalent width equal to the 
equivalent width of the Gaussian component, i.e., 60 eV. Using these data and 
those from basic simulations we have checked how the average depends on the 
relative share of different spectra in entire sample.

\section{Discussion}
\label{discuss}

\begin{figure}
\begin{center}
\includegraphics[width=\columnwidth]{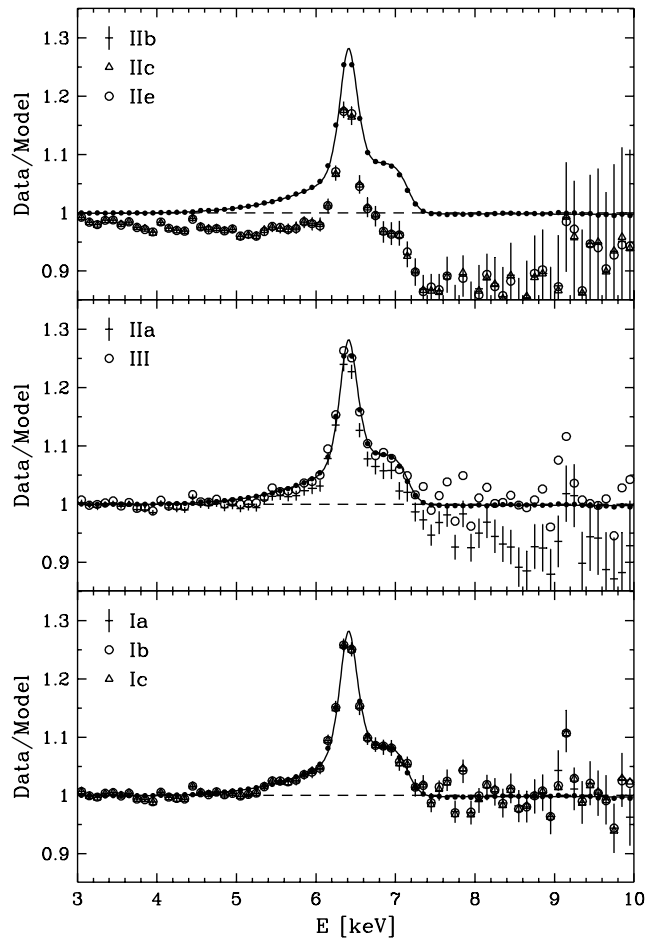}
\end{center}
\caption{\small Averaging methods applied to data/model ratios obtained
for non-binned spectra and continuum model fitted with C-statistic. 
Methods are denoted as in Table \ref{methods}.}
\label{arfig1}
\end{figure}

\begin{figure}
\begin{center}
\includegraphics[width=\columnwidth]{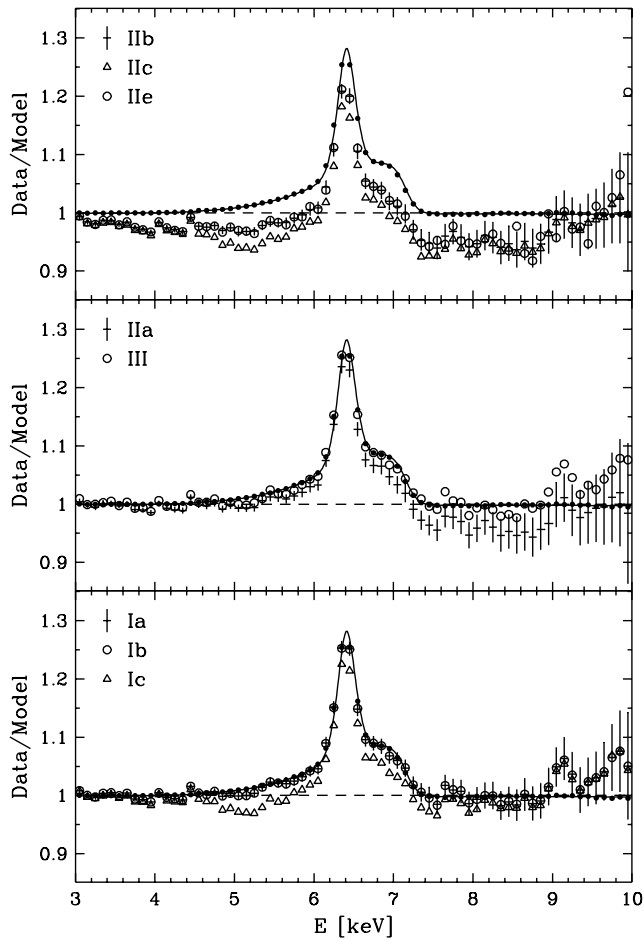}
\end{center}
\caption{\small Averaging methods applied to data/model ratios obtained
for binned up spectra and continuum model fitted with C-statistic. 
}
\label{arfig2}
\end{figure}

\begin{figure}
\begin{center}
\includegraphics[width=\columnwidth]{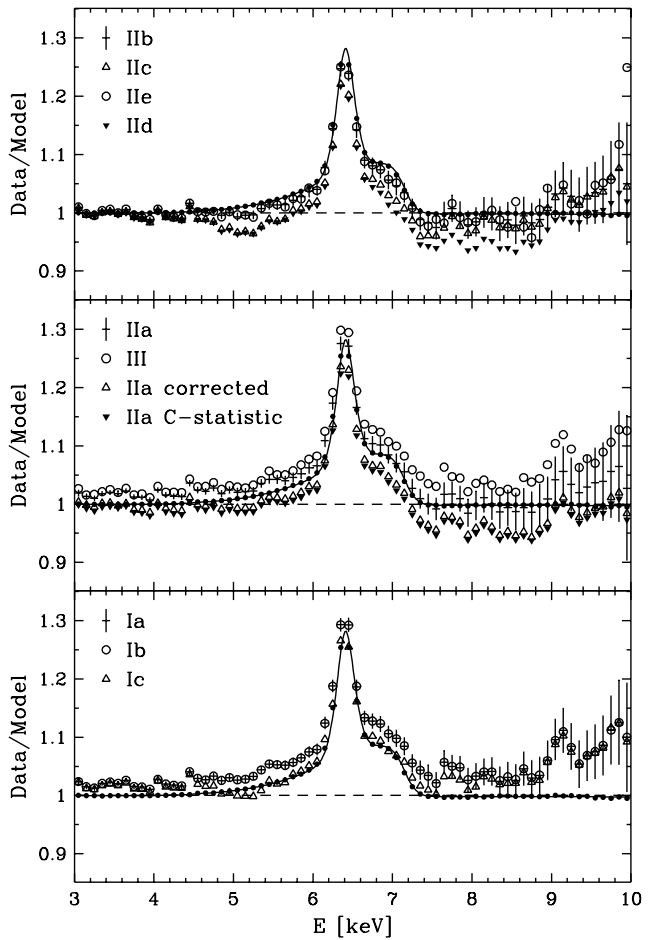}
\end{center}
\caption{\small Averaging methods applied to data/model ratios obtained
for binned up spectra and continuum model fitted with $\chi^{2}$ statistic. 
Correction applied to the results of method IIa is explained in the text,
see Sec. \ref{dibest}.}
\label{arfig3}
\end{figure}

\subsection{Best method}
\label{dibest}

The best results are obtained for methods Ia, Ib and Ic applied to non-binned 
data, as presented in the bottom panel of Fig. \ref{arfig1}. Only these methods
almost perfectly reproduce the shape of the iron line on both its sides. There 
is a spread of results for higher energies, above 7 keV, but these discrepancies 
are symmetric and appear due to small statistics in this energy range. 
Methods Ia and Ib used for data binned up also work well, but here both red and 
blue wings of the line are slightly but systematically underestimated, as can be 
seen from Fig. \ref{arfig2}. For higher energies the spread of results is damped 
due to initial grouping of the input bins, however, owing to the same fact, 
spurious maxima are produced around 9.2 and 9.8 keV. Then, even with the best 
methods, averaging $binned \; up$ data cannot be considered to be reliable for 
the higher energy boundary of $ASCA$ SIS range. The third procedure from this 
group, method Ic, using bin weights $b_{i}$, fails for binned up spectra; this 
behaviour is explained in Sec. \ref{dibiased}.

Methods based on averaging with the standard weights $w_{i}$ (IIa-IIe) 
obviously cannot work properly for a continuum modelled with the C-statistic. 
This is clearly seen in the upper part of Figs. \ref{arfig1} and \ref{arfig2}, 
where all these procedures fail completely to reproduce the continuum slope. 
For the same reason, methods using the probability functions do not work well 
for continuum fitted with the $\chi^{2}$ statistic, this is illustrated in the 
lower part of Fig. \ref{arfig3}. 

The 'standard' method, IIe, applied to binned up data (Fig. \ref{arfig3}), can 
be used as a crude approximation of the average shape. The red and blue wings 
of iron line are here more underestimated than is observed for 
methods Ia-Ic in Fig. \ref{arfig2}. Also the line peak is not well reproduced
and the distortions induced by binning up are present in the high energy end of 
the spectrum. Compared with method IIb, which uses all the input bins 
overlapping with the output bin (with $g_{i}$ correction for boundary overlap), 
the 'standard' method results exhibit a larger spread for higher energies. This 
is simply the consequence of neglecting input bins with centres lying outside 
the output bin. Similarly to method Ic applied to binned up data (bottom part of 
Fig. \ref{arfig2}), method IIc, using bin weights, produces a distorted shape 
for binned up input data (top panel of Figs. \ref{arfig2},\ref{arfig3}). 

Results of two procedures based on prebinning, IIa and III, are shown in
the middle panel of Figs. \ref{arfig1}-\ref{arfig3}. Due to prebinning, method 
IIa is less sensitive to the improper accuracy weighting using $w_{i}$ weights 
than all procedures averaging directly the input ratios (IIb-IIe). This effect
is obviously stronger for single instrumental bins (Fig. \ref{arfig1}), but
appears clearly also for rebinned data (Fig. \ref{arfig2},\ref{arfig3}).

The arithmetic averaging with prebinning, method III, appears to work quite well
for data obtained with the C-statistic, with, however, a larger spread of 
results observed for higher energies in the case of non-binned data (middle 
panel of Fig. \ref{arfig1}). This spread illustrates the effect of neglecting 
any accuracy weights. The overall agreement between arithmetic and weighted 
averages (the proper ones, using the probability function) is understandable, 
due to the fact that the longest and the shortest exposure times used in 
simulations differ only by factor of 2.5, thus the accuracy weights for 
different spectra also do not differ strongly.

The middle part of Fig. \ref{arfig3} shows that the method with prebinning and
standard accuracy weights $w_{i}$, IIa, behaves quite differently from its
counterpart without prebinning, method IIb (upper panel of this Figure). 
This can be explained by two effects. The first is an improper continuum 
model, fitted with $\chi^{2}$ statistics. Using the results presented in rows 
3 and 5 of Table \ref{meanmodels}, we have corrected the average data/model 
ratios with a scaling function equal to the ratio of these mean power law 
models. The middle panel of Fig. \ref{arfig3} presents the results of this 
correction compared to the results of procedure IIa obtained for a continuum 
model fitted with C-statistic (i.e., the same as in the middle panel of Fig. 
\ref{arfig2}). Now these results are in agreement and the rest of difference 
between methods IIb and IIa, applied to '$\chi^{2}$' input data, comes from 
the fact that the latter procedure is less affected by the imperfect accuracy 
weighting by widths $w_{i}$. This effect can be estimated from comparison 
between the results of methods IIb and IIa shown in the upper and middle panels, 
respectively, of Fig. \ref{arfig2}.

\subsection{Biased vs. unbiased binning up}
\label{dibiased}

The average iron line shape for Sy1 nuclei, presented in \citet{LZ2001}, was 
obtained with the procedure IId applied to initially binned up spectra. This 
is the standard manner used in X-ray spectral fitting with a $\chi^{2}$ 
statistic to group single input bins to bins with at least 20 counts. As stated 
in Sec. \ref{biased}, such a binning up is biased. However, it is hard to see 
how important this effect is in the situation where real data of unknown 
average shape are used. A quantitative estimate of the influence of this bias 
on the results of averaging can now be done, for data simulated using a known 
reference model. 

\begin{figure}
\begin{center}
\includegraphics[width=\columnwidth]{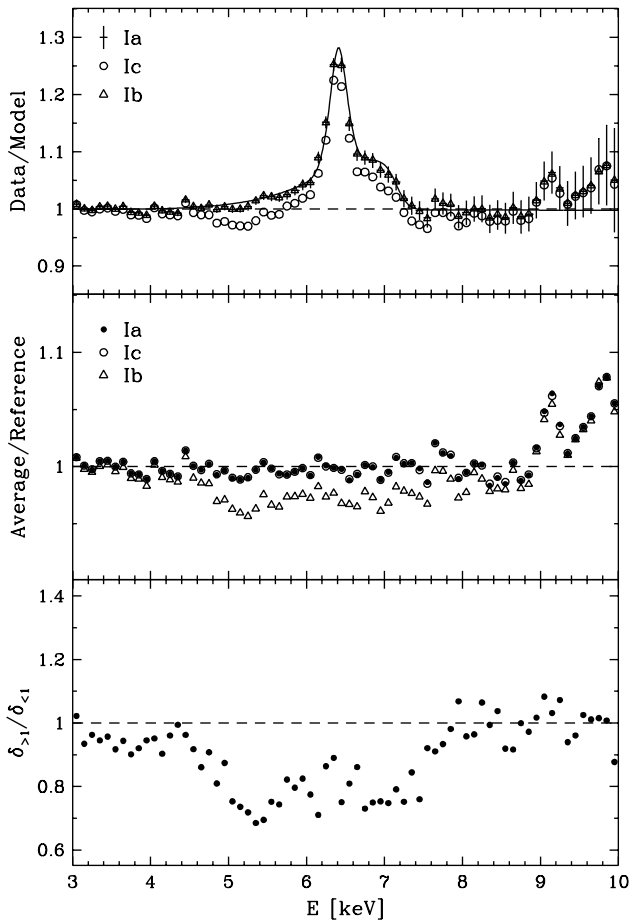}
\end{center}
\caption{\small Test of averaging for biased (correlated) binning up. Upper
panel: averaging results as in the lower panel of Fig. \ref{arfig2}. Middle 
panel shows the ratios between these results and the reference shape. The 
ratios of arithmetic means of the overlap $\delta _{i}$ for data/model ratios 
above and below 1 are presented in the bottom panel.}
\label{arfig6}
\end{figure}

\begin{figure}
\begin{center}
\includegraphics[width=\columnwidth]{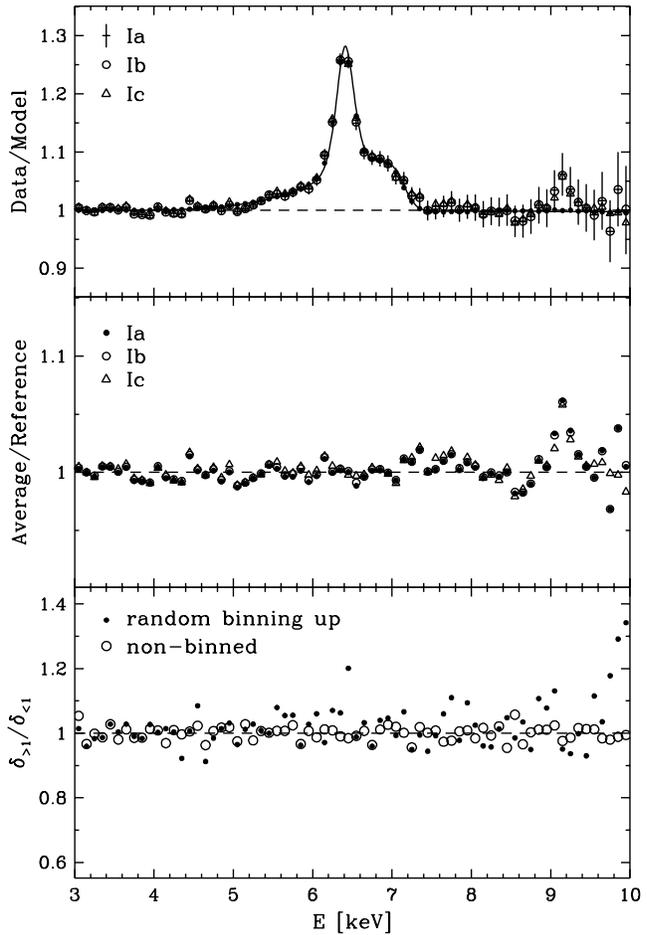}
\end{center}
\caption{\small The same test as in Fig. \protect\ref{arfig6} done for random 
(uncorrelated) binning up. Please note the change of the artificial features
above 9 keV, in comparison with the results shown in Fig. \protect\ref{arfig6},
where different binning up pattern was applied.}
\label{arfig5}
\end{figure}

In the upper panel of Fig. \ref{arfig6} the average spectral shapes obtained 
with methods Ia, Ib and Ic for binned up 'C-statistic' data are presented again 
as in Fig. \ref{arfig2}. To show more clearly the differences between these 
methods, the ratios between their results and the reference average were plotted 
in the middle part of Fig. \ref{arfig6}. The bottom panel of this Figure 
illustrates the ratio of the arithmetic averages of the input and output bins 
overlap, $\delta _{i}$, determined for data/model ratios above 1, 
$\delta _{>1}$, and below 1, $\delta _{<1}$. The deviation between the results 
of method using bin weights, Ib, and the reference results evidently follows 
the departure of $\delta _{>1}/\delta _{<1}$ from unity. Fig. \ref{arfig5} 
presents the results of the same test applied to input data binned up with 
random binning, i.e., where the binning pattern for a given spectrum was 
taken from the other spectrum. Now all tested methods are in concordance, 
the procedure using bin weights reproduces the spectral shape equally well. 
The $\delta _{>1}/\delta _{<1}$ ratios are close to unity, but with a larger 
spread than the same ratios obtained for non-binned data, shown for 
comparison in the bottom part of Fig. \ref{arfig5}. 

The results of method IId, i.e., that employed by \cite{LZ2001}, are presented 
in the upper panel of Fig. \ref{arfig3}, The correction 
introduced in Eq. (\ref{correct}) looks significant for higher energies 
($>$ 6 keV), where $g_{i}$ values are larger. However, as we have tested, the 
differences between the results of methods IIc and IId disappear when these 
procedures are applied to the data binned up in non-biased way. Therefore, 
correction (\ref{correct}) manifests itself mainly by scaling the ratio errors 
to larger, proper values, almost not affecting the ratios themselves.

The scale of discrepancies between the results of procedures using bin weights
(Ic,IIc,IId), applied to the data with biased binning up, and the reference 
results, shown in Figs. \ref{arfig2},\ref{arfig3}, seems to be large. 
Nevertheless, the spectra tested here are extremely weak, and the average 
spectral shape presented in \citet{LZ2001}, obtained for better, on average,
spectra of Sy1 nuclei, is only moderately affected by this mistake. 

\begin{figure}
\begin{center}
\includegraphics[width=\columnwidth]{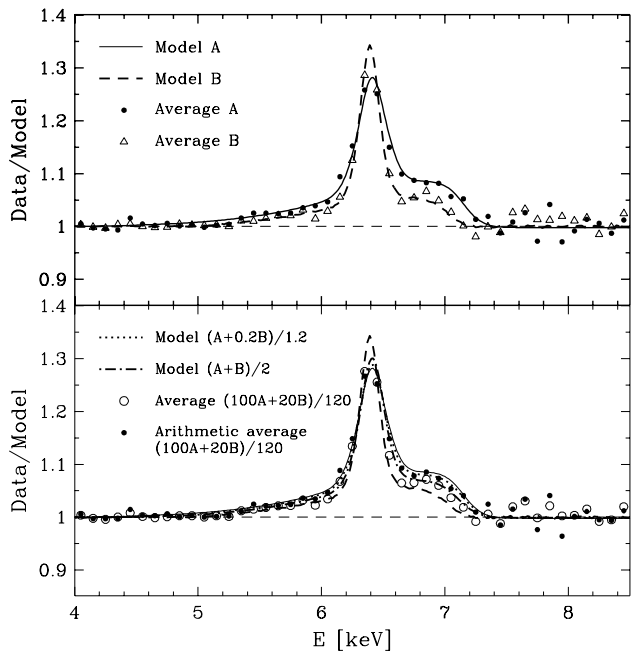}
\end{center}
\caption{\small Test for sample with subsamples corresponding to two distinct 
spectral shapes. Upper panel shows the basic (A) and alternative (B) reference 
models together with averages obtained with method Ia for 100 basic (A) and 20 
alternative (B) spectra. In the lower panel average reference models are
presented, one with weights proportional to the number of spectra in given
subsample, and second, with equal weights. The average for the entire sample
of 120 spectra are performed with methods using the accuracy weights (Ia,
circles) and without it (III, dots).}
\label{arfig4}
\end{figure}

\subsection{Inhomogeneous shapes}
\label{diinhomo}

The results of averaging with method Ia applied to non-homogeneous input data 
are presented in Fig. \ref{arfig4}. Two reference models are clearly distinct;
the blue wings of the disc line component in these models are especially 
different. Assuming that the real composition of the studied objects is in 
proportion 1 to 0.2 to the advantage of the stronger disc line model, we should 
expect the mean to be closer to this model. However, we have assumed (see Sec. 
\ref{inhomo}) that the objects with a weaker disc line component are about 10 
times brighter than the rest of the sample. Then the weighted average 
(circles in the lower panel of Fig. \ref{arfig4}) appears to be much closer to 
the shape of the brighter sources in comparison to the expected mean (dotted 
line in this Figure). This disadvantage can be removed with the use of 
arithmetic average, the results of which are shown with dots in the lower part 
of Fig. \ref{arfig4}. As already mentioned in Sec. \ref{dibest}, arithmetic 
averaging leads to a larger spread in results, thus it is advisable to check 
the homogeneity of entire sample by dividing it to subsamples, grouped 
accidentally or according to some condition. Afterwards, the weighted averages 
of approximately homogeneous subsamples can be averaged arithmetically. 

In the case of Sy1 nuclei the spectra are known to be different for various
objects in various spectral states and usually the values of their physical 
parameters are distributed over a wide range. However, there may exist samples 
of objects exhibiting clearly distinct physical character. Then the average over 
the entire sample will not describe any real object but, again, this can be 
easily checked by comparing the results for subsamples.
                      
\begin{figure}
\begin{center}
\includegraphics[width=\columnwidth]{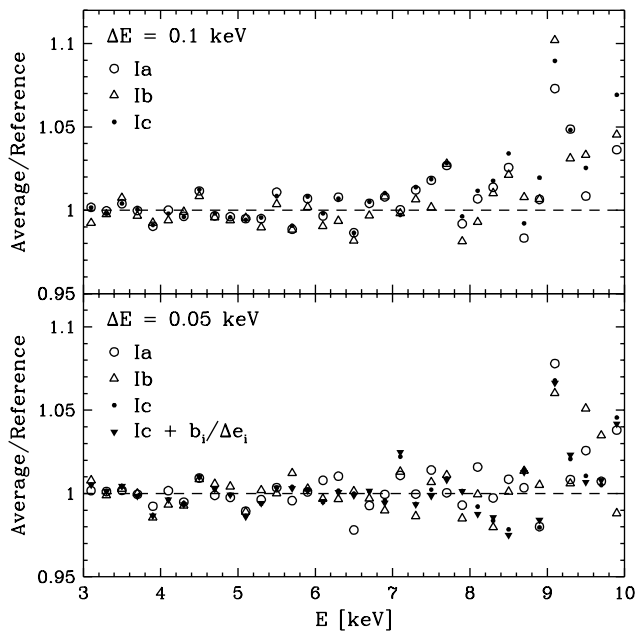}
\end{center}
\caption{\small Averages with and without bin weights. Results of a given
method are presented as the ratio between them and the results of the reference
averaging. The upper panel shows the results for the output bin width equal to
0.1 keV, in the lower panel results for 0.05 keV output bins are presented.
For clarity, results are grouped (arithmetically) to 0.2 keV bins.}
\label{arfig7}
\end{figure}

\subsection{Incomplete information}
\label{diincomplete}

The main reason of using bin weights $b_{i}$ instead of prebinning in 
\cite{LZ2001} was the aim of having the possibility of studying all discrete 
spectral features in the average. There was also a second reason: there were 
many spectra, especially those observed with SIS1 spectrometer, which, due to 
the 'worse quality' flag, were truncated far below the upper instrumental limit (10 keV), at about 
7 or 8 keV. In this case prebinning applied only to completely covered output 
bins leads to the loss of some information. To avoid this, one can treat input 
bins only partially covering the output bin as representative of the whole 
output bin, but now their share to the average is somewhat overestimated. 
On the contrary, usage of bin weights automatically reduces the significance 
of such incomplete information. 

To better study the issue of incomplete information we have applied procedures 
with and without prebinning to the data, where half, chosen randomly, of the 
input bins was neglected in calculating the average. We have tested the best 
methods, i.e., those based on the product of probability density functions, 
using the data with random binning up, since for the non-binned data the 
effect of bin weighting is reduced. The results of these tests are shown in 
Fig. \ref{arfig7}, for two widths of the output bins, 0.1 and 0.05 keV. In the 
case of broader output bins there is no qualitative difference between the 
results of methods Ia and Ib, without bin weights, and method Ic, applying 
$b_{i}$ weights. For narrower output bins, in the region of Fe line peak 
(at 5.5-7 keV) the results of method Ic are more consistent with the reference 
average than the results of method Ib, whereas method Ia still does not seem to 
be clearly worse than method Ic. In general, bin weighting is advisable in the
situation when one is going to study the most discrete features of the spectrum. 
However, for weak spectra the advantage of better resolution may not prevail 
over the benefit of higher accuracy coming from prebinning.

Application of more advanced bin weighting with additional $b_{i}/\Delta 
e_{i}$ weights (see Sec. \ref{rebin}) is also shown in the lower part of Fig. 
\ref{arfig7}, but it does not change the results significantly. Therefore, 
there is no need to apply complex bin weighting for data collected with 
instruments of rather limited spectral resolution.

\vspace{0.4cm}

The simultaneous fitting of many spectra, with the same line shape parameters
but with different continuum models and line normalizations, can be considered
as an alternative to fitting the spectrum obtained from the average shape. 
However, such an approach is inefficient in the case where there are more than 
a few dozen spectra. Handling a large set of input data and a much larger set 
of parameters associated with them is impractical for any fitting software 
due to time consuming computations and problems in assigning accuracy to the 
fitted values. On the other hand, the spectrum with the average shape can be 
easily fitted when the model incorporates some component describing the 
instrumental resolution, e.g., {\it gsmooth} model in XSPEC.

\section{Conclusions}

We have studied the problem of averaging the spectral shapes in the case
of weak X-ray spectra. The method applied in \citet{LZ2001} was substantially 
improved by more correct treatment of the Poisson character data and better 
modelling of the reference continuum. Various alternative approaches used to 
obtain the average spectral shape were tested with simulated data. 

The reference average is correctly reproduced only by the methods based on the 
description of the data uncertainty by probability density functions (Eqs.
17-19). Among them, the method applying prebinning, i.e., summing the number of 
counts within the output bin (Eq. 18), is the simplest. Compared to the 
prebinning method, the procedure incorporating weights associated with the input 
and output bin overlap (Eq. 19) works better only in the case where the output 
bins are narrow and, accordingly, it is recommended for data taken with 
instruments of a very good spectral resolution.

There is no need to initially bin up the averaged data when the methods applying
the probability functions are used and the continuum model is fitted with the
C-statistic. Only in this case is all of the information collected with the 
single, narrow channels of the instrument taken into account. Although the best 
methods also work well for spectra binned up in a non-biased way, the resulting 
spectral shape can be distorted due to the loss or mixing of information 
contained in single bins. 

The usage of the $\chi^{2}$ statistic in modelling of weak spectra leads to 
biased results and deformed average shape. 

The arithmetic average used for prebinned data quite fairly reproduces the 
spectral shape, and can be recommended for approximately equally accurate input 
data or for testing the homogeneity of the averaged sample.

\section*{Acknowledgments}

We would like to thank the referee for all comments and suggestions which
helped to improve the presentation of our results. Discussions with Tomasz 
Bulik, Rafa{\l} Moderski and Grzegorz Wardzi\'nski are warmly acknowledged. 
This research has been supported by grants from KBN (5P03D00821,2P03C00619p1,2).


\appendix

\section{Average flux}
\label{appena}

Energy flux is defined as the ratio of the number of photons $N_{i}$
collected in a given channel $i$ and the product of the energy range of this 
channel $\Delta e_{i}$, exposure time $T$ and effective (for the interesting 
energy range) area $A$ of the instrument

\begin{equation}
\langle f \rangle _{i} = \frac{N_{i}}{AT\Delta e_{i}}.
\end{equation}

Neglecting the relatively small errors of $T$, $\Delta e_{i}$ and $A$ the 
accuracy of the flux is given by

\begin{equation}
\Delta \langle f \rangle _{i} = \frac{\Delta N_{i}}{AT\Delta e_{i}}.
\end{equation}

In the case of no background, for Poisson distributed data, $\Delta N_{i}$ 
is usually approximated by $\sqrt{N_{i}}$, hence the standard weighted 
average of fluxes measured for $n_{k}$ input bins overlapping with the output 
bin has the form

\begin{eqnarray}
\overline{f} _{k} = \frac{\D \sum_{i=1}^{n_{k}} \frac{\D (AT\Delta e_{i})^{2}}
{\D (\sqrt{N_{i}})^2}\frac{\D N_{i}}{\D AT\Delta e_{i}}}
{\D \sum_{i=1}^{n_{k}} \frac{\D (AT\Delta e_{i})^{2}}{\D (\sqrt{N_{i}})^2}}
=\frac{\D AT \sum_{i=1}^{n_{k}} \Delta e_{i}}
{\D (AT)^{2} \sum_{i=1}^{n_{k}} \frac{\D (\Delta e_{i})^{2}}
{\D {N_{i}}}}.
\end{eqnarray}

Again using the equality $N_{i} = \langle f \rangle _{i} \Delta e_{i}AT$ and 
inserting the output bin width $\Delta E_{k}$ in the numerator and denominator 
we obtain

\begin{eqnarray}
\overline{f} _{k} = \frac{\D \sum_{i=1}^{n_{k}} 
\frac{\D \Delta e_{i}}{\Delta E_{k}}}{\D \sum_{i=1}^{n_{k}} 
\frac{\D \Delta e_{i}}{\D \Delta E_{k}} \frac{\D 1}{\D \langle f \rangle _{i}}}.
\end{eqnarray}

This is the harmonic mean, weighted with bin weights. In the situation where
flux is approximately constant within the output bin, the factor $1/\langle f 
\rangle _{i}$ can be taken out of the sum in denominator and we can insert 
$\langle f \rangle _{i}$ in the sum in the numerator. Therefore, for the 
function of flux type, the weighted average given by (\ref{weighted}) behaves 
in some sense similarly to the bin weighted average, Eq.~(\ref{rebinned}). 

The harmonic average is commonly recommended for the quantities of a relative 
character, defined as a ratio. Nevertheless, we cannot test this approach since 
for weak spectra the flux measured for narrow input bins is often equal to 0.

\section{Poisson distribution for unknown background}
\label{appenb}

The X-ray background spectra are often weak and the background rate cannot be 
estimated with good accuracy. This uncertainty obviously affects the accuracy 
of the determined net source counts. There have been proposed solutions to this 
problem, modifying the Poisson distribution used to describe the source + 
background data. Below we compare the results of two of such methods with the 
distribution given by equation (\ref{poiss}). The first procedure, classical 
\citep{RL2001}, is based on the likelihood ratios, the second, Bayesian 
\citep{Loredo1992}, uses marginalization with respect to the background 
rate. For simplicity's sake, we assume here that the source and background 
counts were taken during the same time and for the same size of the signal 
region.

\citet{RL2001} developed their method for setting confidence intervals for 
small signals in the presence of background. To find the confidence region
they use the likelihood ratio test statistic $\Lambda$ in the form

\begin{equation}
    \Lambda(\lambda_{0};x,y) = \frac{max \bigl\{l(\lambda_{0},\eta;N,B):
    \eta\geq 0\bigr\}}
    {max \bigl\{l(\lambda,\eta;N,B):\lambda\geq 0, \eta\geq 0\bigr\}},
\label{liketest}
\end{equation}

where $\lambda_{0}$ is the tested null hypothesis value of the net source counts 
$\lambda$. The likelihood function for $\lambda$ source and $\eta$ background 
counts, given the observed $N$ source counts and $B$ background counts, 
is expressed as the product of two Poisson distributions

\begin{equation}
    l(\lambda,\eta;N,B) = \frac{(\lambda+\eta)^{N}}{N!}e^{-(\lambda+\eta)}
    \frac{\eta^{B}}{B!}e^{-\eta}.
\label{likefun}
\end{equation}

As can be expected, the probability distribution derived from the likelihood 
ratio test is broader than the distribution obtained for the case of fixed
background (Eq. (\ref{poiss})). This is the effect of calculating the likelihood
function with variable $\eta$, what corresponds to taking the background 
uncertainty into account in this method.

\begin{figure}
\begin{center}
\includegraphics[width=\columnwidth]{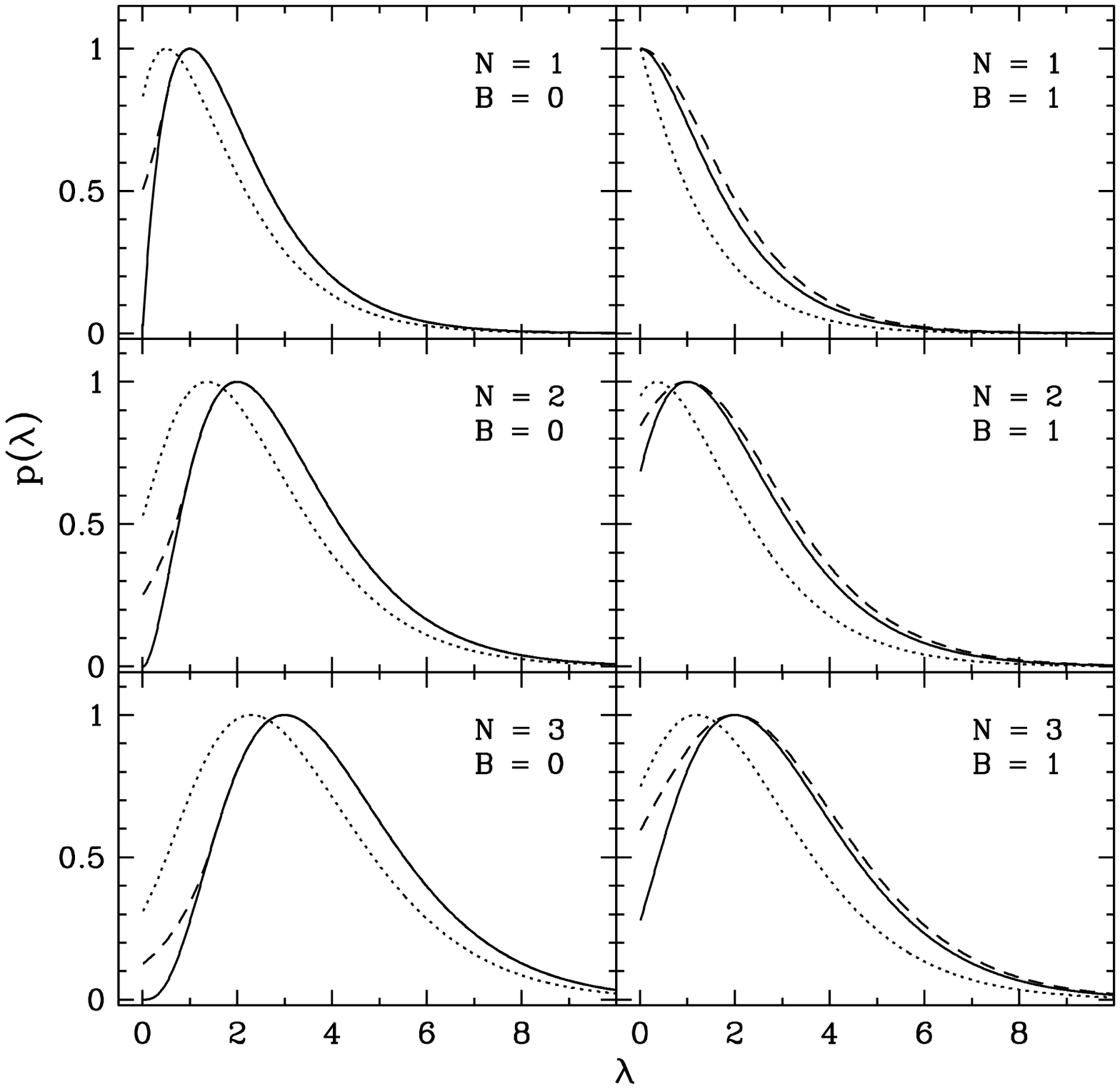}
\end{center}
\caption{\small Probability distributions for net source counts $\lambda$ 
obtained with the use of different methods. N and B are the measured source 
and background counts, respectively. Solid line shows distribution obtained 
for the case, when the background rate is assumed to be known (Eq. 
(\ref{poiss})). Dashed line presents the distribution based on the likelihood 
ratio test, Eq.~(\ref{liketest}). Bayesian probability distribution, derived 
from the marginalization with respect to the background rate, Eq. (\ref{margi}), 
is plotted with the dotted line.}
\label{polodis}
\end{figure}

The shape of the likelihood ratio $\Lambda(\lambda)$, normalized to obtain 
maximal value equal to 1, is shown in Fig. \ref{polodis}. In the case, when the 
measured background is equal to 0, broadening affects the distribution only 
in the lowest, close to zero, $\lambda$ region. (For $B$=0 we have modified the
Rolke \& Lop\'ez formula for $\eta$ maximizing the numerator in Eq. 
(\ref{liketest}), using the relation $\eta_{max}=max(0,(N-2\lambda_{0})/2)$.) 
The mode for this distribution is equal to $N$, as in the case of distribution 
obtained for the known background case. However, the mean for $\Lambda
(\lambda)$ distribution is smaller or larger than $N+1$, depending on $N$ 
and $B$.

The Bayesian method of inferring the signal strength in the situation of 
imprecisely measured background is presented in \citet{Loredo1992}. First we 
quote the formula for the posterior probability density derived by Loredo for 
the case when the background counts number $\eta$ is known

\begin{equation}
    p(\lambda;N,\eta) = C\frac{(\lambda+\eta)^{N}e^{-(\lambda+\eta)}}{N!},
\label{appois}
\end{equation}

where the normalization constant $C$ is equal to $\big\{\sum_{i=0}^{N}\eta^{i}
e^{-\eta}/i!\big\}^{-1}$. The mode for this distribution is equal to $N$ and 
the mean, $\overline\lambda$, equals to $N+1$. 

For the unknown background case, the nuisance parameter, background rate $\eta$, 
is eliminated through marginalization, i.e., through finding the integral of
Eq. (\ref{appois}) over $\eta$. The posterior probability distribution is 
calculated using the expansion of the binomial $(\lambda+\eta)^{N}$ and has 
the form

\begin{equation}
    p(\lambda;N,B) = \sum_{i=0}^{N}C_{i}\frac{\lambda^{i}e^{-\lambda}}{i!},
\label{margi}
\end{equation}

with coefficients $C_{i}$ given by formula

\begin{equation}
    C_{i} = \frac{2^{i}\frac{(N+B-i)!}{(N-i)!}}
    {\sum_{j=0}^{N}2^{j}\frac{(N+B-j)!}{(N-j)!}}.
\end{equation}

The above formula is interpreted in terms of a weighted average of the
posterior densities calculated attributing 0, 1, ..., N events to the signal.
The $C_{i}$ weights are the probabilities that $i$ of the events observed 
on-source are from the source, provided that B counts are measured off-source.  

The Bayesian probability distribution, computed from equation (\ref{margi}) and 
normalized to have maximum equal to 1, differs clearly from two other 
distributions shown in Fig. \ref{polodis}. Its right tail is more extended but 
the major difference is the shift of its maximum towards lower values of 
$\lambda$. Depending on how N compares to B, this shift correlates inversely 
with the measured background counts, exceeding one unit for $B=0$ and 
approaching 0 for $B=N$. Using the relation $\int_{0}^{\infty}\lambda^{i}
e^{-\lambda}d\lambda = i!$ it is easy to show that the mean $\overline \lambda$ 
for the probability density given by (\ref{margi}) is equal to $\sum_{i=0}^{N}
C_{i}(i+1)/\sum_{i=0}^{N}C_{i}$. Hence, taking into account the background 
uncertainty in the Bayesian procedure leads to lowering of the mean value, with 
the decrease depending on the measured background counts.

\bsp

\label{lastpage}

\end{document}